\documentclass[conference,a4paper]{APSIPA2023}
\usepackage{multirow}
\usepackage{amsmath}
\usepackage[psamsfonts]{amssymb}
\usepackage{amsxtra}
\usepackage{threeparttable}
\usepackage{graphicx}
\usepackage{booktabs}
\usepackage{float}
\restylefloat{table}
\usepackage{array}
\newcolumntype{C}{>{\centering\arraybackslash}p{3.5em}}
\usepackage{float}
\floatstyle{plaintop}
\restylefloat{table}

\usepackage{geometry}
\usepackage{fancyhdr}

\fancypagestyle{firststyle} {
    \fancyhf{}
    \fancyhead[C]{2023 Asia Pacific Signal and Information Processing Association Annual Summit and Conference (APSIPA ASC)}

}

\begin{document}

\title{Analysis of Speech Separation Performance Degradation on Emotional Speech Mixtures}

\author{%
\authorblockN{%
Jia Qi Yip\authorrefmark{1}\authorrefmark{2}, Dianwen Ng\authorrefmark{1}\authorrefmark{2}, Bin Ma\authorrefmark{1}, Chng Eng Siong\authorrefmark{2}
}
\authorblockA{%
\authorrefmark{1}
Speech Lab of DAMO Academy, Alibaba Group \\
}
\authorblockA{%
\authorrefmark{2}
Nanyang Technological University, Singapore\\
E-mail: jiaqi006@e.ntu.edu.sg
}
}

\maketitle
\thispagestyle{firststyle}
\pagestyle{fancy}

\begin{abstract}
Despite recent strides made in Speech Separation, most models are trained on datasets with neutral emotions. Emotional speech has been known to degrade performance of models in a variety of speech tasks, which reduces the effectiveness of these models when deployed in real-world scenarios. In this paper we perform analysis to differentiate the performance degradation arising from the emotions in speech from the impact of out-of-domain inference. This is measured using a carefully designed test dataset, Emo2Mix, consisting of balanced data across all emotional combinations. We show that even models with strong out-of-domain performance such as Sepformer can still suffer significant degradation of up to 5.1 dB SI-SDRi on mixtures with strong emotions. This demonstrates the importance of accounting for emotions in real-world speech separation applications.
\end{abstract}






\section{Introduction}
Speech Separation is the task of obtaining single speaker speech from a mixture of speakers, also known as the cocktail party problem~\cite{cherry1953}. It has been the focus of much recent research, as many downstream speech models for tasks such as Automatic Speech Recognition (ASR)~\cite{ng2023dehubert} are trained on single talker speech. For deployment, mixed speech received from the wild should ideally be separated before performing ASR. 

The effect of a speaker's emotions on their speech~\cite{Chakraborty2018AnalyzingEI} and speaker emotion recognition~\cite{AKCAY202056} are well studied research areas. The emotions of a speaker or the intention of a speaker to convey emotion can result in differences in speech. While some of the differences are semantic~\cite{johnstone_2017}, there are clear differences also manifest in the prosody~\cite{SCHERER2003227} and articulation~\cite{Lee2005AnAS} of words. Thus, emotional differences in speech can show up at the frame level, where speech separation models operate. 

The difference in sound between neutral or read speech and emotional speech has been shown to lead to performance degradation on a number of speech tasks such as speaker verification~\cite{pappagari2020xvectors} and ASR~\cite{munot-nenkova-2019-emotion}. Conversely, a mixture of speakers with different emotions also results in a degradation of emotion recognition performance~\cite{9882231}.

\begin{figure}[t]
  \centering
  \includegraphics[width=0.4\textwidth]{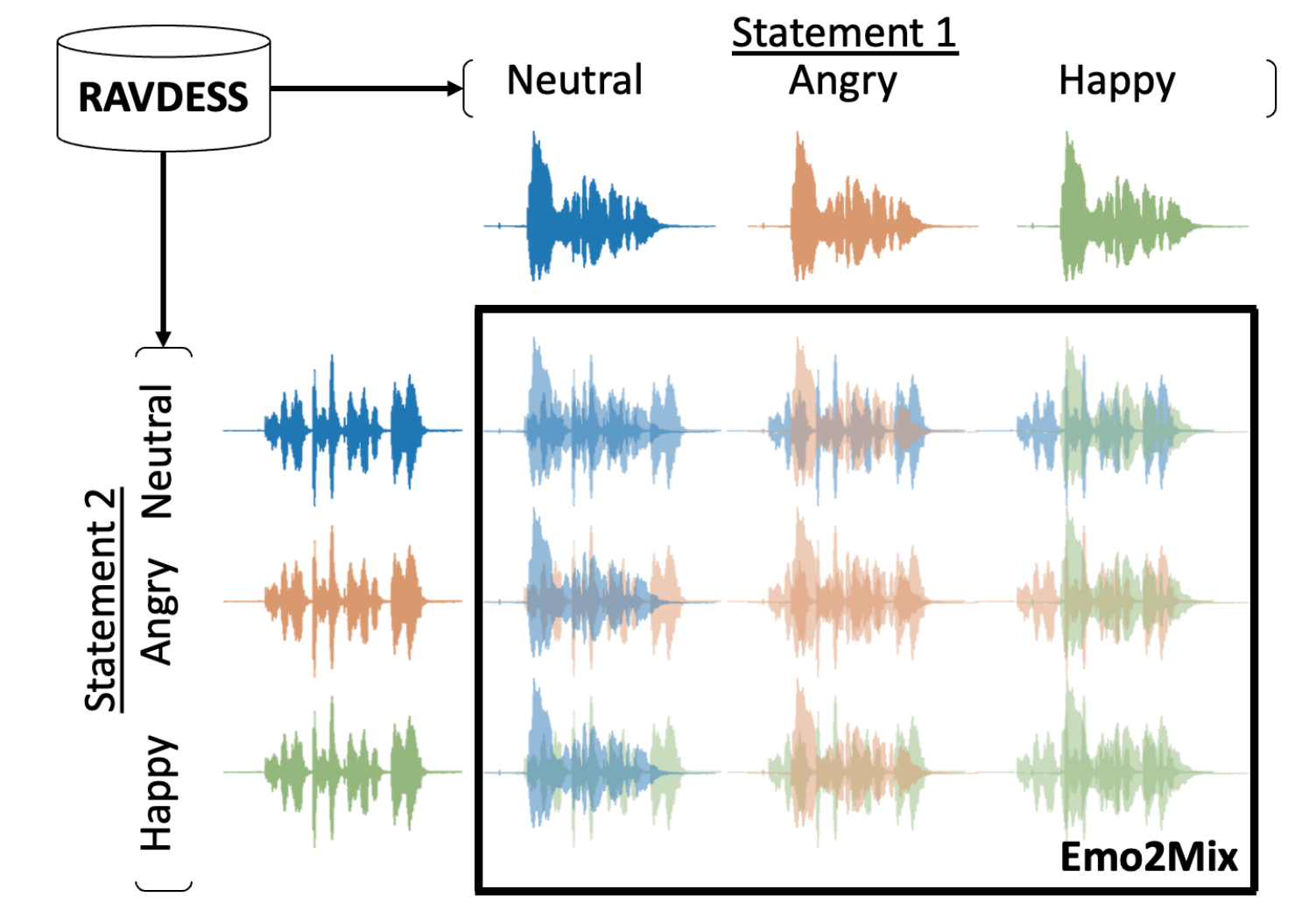}
  \caption{Illustration of the mixing strategy. Only three emotions are shown here for brevity but the full set of available emotions are used in the dataset (8 at normal intensity and 7 at strong intensity). Each mixture consist of different speakers each vocalizing different statements. The same two statements are shared across emotions and speakers to control for semantic differences.}
  \label{fig:emo2mix_illustration}
\end{figure}

The separation of emotional speech is important because overlapping speech often occurs when speakers are at heightened emotional states, such as in a heated argument or excited conversation, where the typical decorum of turn taking is breached~\cite{Gervits2018TowardsAC}. Speech Separation models are often trained on datasets built from neutral speech. The most common datasets used for speech separation are LibriMix~\cite{librimix} created from the LibriSpeech~\cite{librispeech} corpus and WSJ0Mix~\cite{wsj0-2mix} created from the WSJ0~\cite{wsj0} corpus. Both datasets consist of read speech recorded under controlled conditions and neither are explicitly emotional datasets. This results in a potential mismatch between speech separation models trained primarily on neutral speech and the real-world emotional speech mixtures.

\subsection{Our Approach}
In this paper analyses the performance degradation of speech separation models using a custom speech separation test dataset, Emo2Mix, built based utterances from the Ryerson Audio-Visual Database of Emotional Speech and Song (RAVDESS)~\cite{ravdess}. To the authors' knowledge this is the first study to properly investigate the impact of emotions on speech separation with a test dataset that is balanced across multiple emotions at two different emotional intensities. 

Using this dataset, we conduct a detailed analysis of the impact of emotional speech on state-of-the-art speech separation models that were trained on neutral speech. To control for the out-of-domain issue due to the unseen speakers in the Emo2Mix test dataset, we make use of the neural emotion utterances present within the RAVDESS dataset as a baseline for an out-of-domain speaker with an ``in-domain" emotion.

We show that Emo2Mix presents a significant challenge to speech separation models trained on neutral speech with a recent state-of-the-art Sepformer model~\cite{sepformer1}\cite{sepformer2} experiencing a performance degradation of  7.0dB compared to its Libri2Mix baseline. Of this, a performance difference of 1.9dB can be attributed to emotions alone, when comparing against the in-domain neutral emotion baseline.

\subsection{Related Work}
Recently, ~\cite{ravdess2mix} also proposed an emotional mixture test dataset based on the RAVDESS~\cite{ravdess} dataset, RAVDESS2Mix. The dataset consists of mixtures from the RAVDESS dataset and LibriSpeech~\cite{librispeech} dataset, as well as enrollment speech from the RAVDESS dataset for the target speaker extraction task.  

Compared to RAVDESS2Mix~\cite{ravdess2mix}, our proposed Emo2Mix dataset does not consist of any utterances from the LibriSpeech corpus. This is for two reasons. Firstly, the LibriSpeech corpus consists exclusively of neutral speech, which limits the number of emotion combinations possible for the dataset. Secondly, many speech separation models are trained on the Libri2Mix dataset, which uses utterances from the LibriSpeech corpus. This makes the RAVDESS2Mix test dataset biased, as models pretrained on Libri2Mix will have and advantage over models pretrained on other datasets such as the commonly used WSJ02Mix dataset, which creates utterances from the WSJ0 corpus. Based on the RAVDESS2Mix ~\cite{ravdess2mix} proposed that target speaker extraction is negatively impacted by emotional speech while blind source separation was not as significantly impacted. However, using Emo2Mix, we can show that the same performance degradation is observed once the in-domain advantage of Librispeech models are removed.   

\section{Methodology}
\subsection{Evaluation}
The performance of the models are measured using Scale-invariant signal-to-distortion ratio improvement (SI-SDRi) which is measured by comparing the output waveforms of the models with the ground-truth waveforms. Since there are multiple outputs, the final performance is calculated using the permutation invariant approach.
\subsection{The RAVDESS dataset}
The RAVDESS dataset\cite{ravdess} is an emotional dataset consisting of 7356 recordings by 24 actors, gender balanced with 12 male and 12 female. The actors vocalize two lexically-matched statements in a neutral North American accent at two emotional intensities, normal and strong. At the normal emotional intensity, 8 emotions are expressed (neutral, calm, happy, sad, angry, fearful, disgust, surprised) while the strong intensity share the same emotions except for neutral. 

RAVDESS has a number of advantages over other emotional speech databases because of its inclusion of two neutral emotions, ``neutral" and ``calm" and two emotional intensities ``normal" and ``strong". These features are useful for comparison with existing corpora like LibriMix and WSJ0Mix which also call in this emotional category. These baselines can help differentiate between the performance impact of out-of-domain inference from the unseen speakers and the actual performance impact of the emotional speech.

\subsection{Emo2Mix}
The Emo2Mix dataset draws from the RAVDESS dataset to produce two-speaker mixtures of emotional speech. The subset of RAVDESS selected for Emo2Mix excludes singing speech. It also excludes the audiovisual data. A breakdown of the features in the chosen subset of the RAVDESS dataset is shown in Table~\ref{tab:ravdess_breakdown}.

\begin{table}[h!]
  \caption{Breakdown of the features of the selected subset of the RAVDESS dataset used to create Emo2Mix}
  \vspace{5pt}
  \label{tab:ravdess_breakdown}
  \centering
  \begin{tabular}{lcc}
    \toprule
    \textbf{Feature} & \textbf{Normal} & \textbf{Strong} \\
    \midrule
    Number of Speakers & 24 & 24 \\
    Number of Emotions & 8 & 7 \\
    Number of Statements & 2 & 2 \\ 
    Number of Repetitions & 2 & 2\\
    \midrule
    Total Recordings & 768 & 672 \\
    \bottomrule
  \end{tabular}
\end{table}

\begin{table*}[h!]
\caption{Overall results across different datasets, with aggregated results reported for the emotional datasets. The performance degradation due to only emotional mixtures is calculated by (4)-(6) which is the difference in performance between the Strong emotion and the Neutral emotion baseline. Out-of-domain performance degradation, independent of emotions can be measured by the difference between the original baseline and the Emo2Mix (Neutral) mixtures, calculated by (1)-(4).}
\vspace{5pt}
\label{tab:overall_results}
\large\centering
\begin{tabular}{|c|c||c|c|c|c||c|c|c|c|}
\hline
 & \multicolumn{1}{r||}{Model:} &
  \multicolumn{4}{c||}{Sepformer} &
  \multicolumn{4}{c|}{ConvTasNet} \\ \hline
 & \multicolumn{1}{r||}{Training Dataset:} &
  \multicolumn{2}{c|}{Libri2Mix} &
  \multicolumn{2}{c||}{WSJ0-2Mix} &
  \multicolumn{2}{c|}{Libri2Mix} &
  \multicolumn{2}{c|}{WSJ0-2Mix} \\ \hline\hline
\textbf{ID} & \textbf{Testing Dataset} &
  \multicolumn{1}{c|}{\textbf{SI-SDRi}} &
  \multicolumn{1}{c|}{\textbf{SDRi}} &
  \multicolumn{1}{c|}{\textbf{SI-SDRi}} &
  \textbf{SDRi} &
  \multicolumn{1}{c|}{\textbf{SI-SDRi}} &
  \multicolumn{1}{c|}{\textbf{SDRi}} &
  \multicolumn{1}{c|}{\textbf{SI-SDRi}} &
  \textbf{SDRi} \\ \hline
1 & Libri2Mix-Test      & 20.6 & 20.9 & 17.0 & 17.5 & 15.1 & 15.5 & 10.0 & 10.6 \\ \hline
2 & RAVDESS2Mix-Normal  & 20.2 & 20.7 & 15.1 & 16.0 & 15.0 & 15.5 &  8.6 &  9.4 \\ \hline
3 & RAVDESS2Mix-Strong  & 20.2 & 20.7 & 14.4 & 15.3 & 14.9 & 15.5 &  8.2 &  9.1 \\ \hline
4 & Emo2Mix (Neutral)   & 15.5 & 16.3 & 12.0 & 12.9 & 10.5 & 11.4 &  4.9 &  5.8 \\ \hline
5 & Emo2Mix-Normal      & 15.4 & 16.2 & 11.4 & 12.2 & 10.7 & 11.4 &  3.3 &  4.3 \\ \hline
6 & Emo2Mix-Strong      & 13.6 & 14.7 &  9.4 & 10.3 &  8.4 &  9.2 &  0.9 &  2.7 \\ \toprule\bottomrule
7 & (1) - (4)           &  \textbf{5.1} &  \textbf{4.6} &  \textbf{5.0} &  \textbf{4.6} & 4.6  & 4.1  & 5.1  & 4.8  \\ \hline
8 & (4) - (6)           &  \textbf{1.9} &  \textbf{1.6} &  \textbf{2.6} &  \textbf{2.6} &  2.1 &  2.2 &  4.0 &  3.1 \\ \hline
\end{tabular}
\vspace{-10pt}
\end{table*}

The mixtures are created with an even distribution across 64 combinations of 8 emotions at a normal emotional intensity and 49 combinations of 7 emotions at a strong emotion intensity. Each statement in the mixture is different, so the mixtures are produced as permutations and not combinations of emotions. While each actor will vocalize the same statement at the same emotion twice, to reduce superfluous mixtures we simply randomly select one of the two repetitions. 

For each of the 113 emotion-intensity combinations, each actor vocalizes 2 different statements, from which we can create $N\times(N-1)$ combinations can be obtained, since we cannot mix a actor with themselves. For the full dataset, $N=24$, which results in 552 mixtures per emotion-intensity permutation which would result in $552\times113 = 62,376$ mixtures overall. To reduce the number of testing mixtures, we reduce the number of speakers in the test set by setting $N=8$, which results in 56 mixtures per emotion-intensity permutation. This gives $56\times64=3,584$ mixtures for the normal intensity test set, and $56\times49=2,744$ mixtures for the strong intensity test set. This is done by selecting only every third speaker (1,4,7,10,13,17,21,24) from the dataset.
This approach, as opposed to randomly selecting 8 out of 24 of the speakers, is chosen so as to enable the possibility of future work on fine-tuning-models on the other 16 held-out speakers. 

The utterances selected from the RAVDESS dataset are dynamically mixed~\cite{wavesplit} so that the mixture test set does not have to be separately stored. During mixing, a number of pre-processing steps are performed in line with the parameters used for dynamic mixing~\cite{wavesplit} implemented on the Speechbrain~\cite{speechbrain} framework. Two audio segments are first cropped to the length of the minimum of the two, then downsampled to 8kHz and normalized. These processed segments are then combined at equal weights if no clipping is observed, or reweighted if clipping is present. To ensure the reputability of the results we used fixed random seeds for all dataset generation and provide a standard script for generation which we release on GitHub\footnote{https://github.com/Yip-Jia-Qi/EmoMix}. 

\subsection{Training Datasets}
In this work we focus on models that have been trained on the Libri2Mix~\cite{librimix} and WSJ0-2Mix~\cite{wsj0-2mix} datasets. The Libri2Mix training set consists of 50,800 mixtures while WSJ0-2Mix training set consists of 20,000 mixtures. Both have test sets of 3,000 mixtures. We note that 3,000 mixtures are comparable to the 3,584 and 2,744 mixtures in the Emo2Mix test set.

\subsection{Baseline Models}
We make use to two baseline models, Sepformer which adopts a dual-path speech separation architecture with transformer blocks, and ConvTasNet~\cite{convtasnet} which utilizes temporal convolution blocks. Both models are seminal in the field of speech separation, with Sepformer~\cite{sepformer1}~\cite{sepformer2} being the more recent state-of-the art model. We implement the two models on the Speechbrain framework as per their respective papers. For Sepformer we make use of pre-trained checkpoints available from the speechbrain huggingface\footnote{https://huggingface.co/speechbrain}. Two versions of the model are available, one trained on Libri2Mix (Sepfromer-L2M) and one trained on WSJ0-2Mix (Sepformer-WJ2). For ConvTasNet we train the model from scratch and validate the model replicating the Libri2Mix-Test baseline. To align with the models available on hugging face, we also train the model on both the Libri2Mix (ConvTasNet-L2M) and WSJ0-2Mix (ConvTasNet-WJ2) datasets. Although newer models surpassing Sepformer have recently been released~\cite{mossformer}\cite{tfgridnet}, Sepformer and ConvTasNet are currently still more widely used and replicated and thus serve as better baselines.

\begin{table*}[ht!]
\caption{Performance of Sepformer-L2M on the Emo2Mix-Normal dataset broken down by emotion combination reported using SI-SDRi.}
\label{tab:sep-l2m-nrml}

\begin{tabular}{|l||c||*{8}{C|}}
\hline
\\[-1em]
 & RD2M & calm & happy & sad & angry & fearful & disgust & surprised & neutral
\\\hline
\\[-1em]
calm        & \textbf{20.6} & 16.0 & 15.6 & 16.0 & \textbf{17.1} & 15.0 & 14.9 & \underline{14.6} & 15.9 \\
happy       & 20.3 & 15.7 & 14.9 & 16.3 & 16.3 & 15.9 & 16.0 & \underline{13.5} & \textbf{16.4} \\
sad         & 20.4 & \underline{15.6} & 15.8 & 16.1 & \textbf{17.1} & 16.1 & 15.7 & 15.7 & 16.3 \\
angry       & 20.3 & 16.5 & 14.9 & \textbf{17.2} & 15.0 & 16.0 & 15.6 & \underline{14.4} & 16.9 \\
fearful     & 20.2 & 15.4 & 14.3 & \textbf{16.9} & 15.1 & 15.9 & 15.0 & \underline{13.6} & 16.4 \\
disgust     & 20.0 & 15.4 & 15.0 & \textbf{16.3} & 15.3 & 15.5 & 15.1 & \underline{13.9} & 14.7 \\
surprised   & \underline{19.7} & 14.4 & 12.7 & \textbf{15.3 }& 15.2 & 14.5 & 13.8 & \underline{12.5} & 14.5 \\
neutral     & 19.9 & 15.8 & 15.8 & 16.3 & 16.2 & \textbf{16.8} & 15.9 & \underline{14.8} & 15.5 \\
\hline
\end{tabular}
\vspace{-5pt}
\end{table*}

\begin{table*}[ht!]
\caption{Performance of Sepformer-L2M on the Emo2Mix-Strong dataset broken down by emotion combination reported using SI-SDRi.}
\label{tab:sep-l2m-strg}
\begin{tabular}{|l||c||*{7}{C|}}
\hline
\\[-1em]
& RD2M & calm & happy & sad & angry & fearful & disgust & surprised
\\\hline
\\[-1em]
calm          & 20.4 & 15.9 & 17.1 & \underline{15.8} & \textbf{18.6} & 17.7 & 16.9 & 15.9 \\
happy       & \textbf{20.6} & \textbf{17.2 }& 12.4 & 13.8 & \underline{10.9} & 11.2 & 14.8 & 12.2 \\
sad            & 20.1 & 15.7 & 13.6 & 14.7 & \textbf{15.8} & \underline{12.5} & 13.1 & 13.0 \\
angry        & 20.4 & \textbf{17.6} & 11.6 & 15.9 & \underline{9.3} & 10.6 & 15.7 & 13.6 \\
fearful      & 20.3 & \textbf{18.4} & 11.9 & 14.1 & 11.6 & \underline{9.7} & 14.9 & 11.7 \\
disgust      & 19.8 & \textbf{17.4} & 11.7 & 14.4 & 13.9 & 11.8 & 13.4 & \underline{10.7} \\
surprised & \underline{19.7} & \textbf{16.7} & 12.4 & 14.3 & 13.0 & 11.9 & 12.4 & \underline{10.4} \\
\hline
\end{tabular}
\vspace{-5pt}
\end{table*}

\section{Results}
In Table~\ref{tab:overall_results} we compare the overall separation performance of the baseline models against three different datasets, the original Libri2Mix Test dataset, RAVDESS2Mix and Emo2Mix. For each of the baseline models, we report the results based on pre-training on two different sets of training data, Libri2Mix and WSJ02Mix. The results on the RAVDESS2Mix dataset are provided as a point of comparison between related emotional speech separation test sets. 

\subsection{Performance degradation due to emotional speech}
The results presented in Table~\ref{tab:overall_results} for Emo2Mix-Normal and Emo2Mix-Strong are aggregated across all emotional combinations through simple averaging, since all emotional combinations are equally weighted. The Emo2Mix (Neutral) test is a subset of the Emo2Mix-Normal test data and consists of 56 mixtures of different speakers where both speakers are portraying neutral emotions. The purpose of Emo2Mix (Neutral) is to serve as a baseline where the speaker is out-of-domain, but has a neutral emotion matching the emotions in the training data. This allows us to measure the out-of-domain performance degradation, as per (7) in Table~\ref{tab:overall_results}, independent of the impact of emotions. 

Any further performance degradation beyond the Emo2Mix (Neutral) baseline, i.e. (8) in Table~\ref{tab:overall_results}, is unlikely to be due to the out-of-domain inference. If emotions do not impact speech separation performance, we should see the values in row (8) approach zero. However, the results of our experiments in Table~\ref{tab:overall_results} show that this is not the case, as the values for (8) range between 1.6dB and 4.0dB. Thus, based on the comparisons of the out-of-domain and emotional degredation in (7) and (8) of Table~\ref{tab:overall_results}, we can conclude that emotions do result in performance degradation in speech separation models. However, the performance impact of performing inference on an unseen speaker is greater than the performance impact from the within-speaker variance resulting from emotions. 

Taking  (8) divided by (7), we see across multiple models that strong emotions can cause an additional 35-55\% performance drop on top of the out-of-domain performance drop (7) when compared to the in-domain baseline (1). This shows that although the performance degradation due to emotions (8) is less than that from out-of-domain inference (7), the impact is still significant.

We can further support the above conclusion using at the internal consistency in the results of the Emo2Mix test datasets at different emotional intensities. From the results of (4), (5) and (6) in Table~\ref{tab:overall_results} we see a significant performance gap between Emo2Mix-Normal and Emo2Mix-Strong, while the gap between Emo2Mix (Neutral) and Emo2Mix-Normal is much smaller. This suggests that emotional speech causes a deterioration in speech separation performance. If emotions do not impact speech separation performance, we should see that Emo2Mix-Normal and Emo2Mix-Strong have the similar performance. Since this is also a within-dataset comparison, the performance difference is not due to differences in speakers, since the speakers in both Emo2Mix-Normal and Emo2Mix-Strong are identical.

\subsection{Comparison with RAVDESS2Mix}
The RAVDESS2Mix dataset consists of mixtures of 2 speakers where one speaker is drawn from the RAVDESS dataset while the other speaker is drawn from the Libri2Mix-Test dataset. This means that for a model trained on the Libri2Mix dataset, one of the speakers would be in-domain while the other is out-of-domain. In a 2-speaker separation problem, knowing the mask for one speaker could allow a model to infer the mask for the other speaker. This gives a model trained on the Libri2Mix dataset a big advantage. This is observed across all models, where we see little to no performance degradation for Sepformer-L2M and ConvTasNet-L2M but a significant performance drop for Sepformer-WJ2 and ConvTasNet-WJ2. 

Furthermore, when comparing the performance of RAVDESS2Mix-Normal and RAVDESS2Mix-Strong, the latter test set results in a lower performance for models trained on the WSJ0-2Mix dataset but this is not observed for the models trained on the Libri2Mix dataset. This lack of internal consistency across models trained on different training datasets, points to the fact that the use of the Libri2Mix-Test set in the RAVDESS2Mix mixtures allow for the model to bypass having to recognise emotional speech by inferring from the neutral speech from the in-domain Libri2Mix-Test utterance.

In comparison to RAVDESS2Mix, Emo2Mix does not suffer from such an issue and is able to accurately benchmark the impact to performance caused by emotional speech. Across all the models tested, for both Libri2Mix and WSJ0-2Mix trained versions, there is a significant performance difference. This shows that the test dataset we have developed for our analysis, Emo2Mix, is a better benchmark for emotional speech separation than the previously reported RAVDESS2Mix.

\subsection{The benefits of a larger training dataset}
It is common to find in various speech domains such as ASR~\cite{ng2023dehubert} and Keyword spotting~\cite{mixup}~\cite{i2cr}, that a larger training dataset results in better performance. Since the Libri2Mix training dataset is significantly larger than WSJ0-2Mix we would expect to see a similar trend. Furthermore, Libri2Mix consists of LibriSpeech utterances which come from audiobooks which may be closer in domain to the RAVDESS datset. That said, the larger size of the Libri2Mix dataset likely improves the performance of the model on Emo2Mix. Looking only at the Sepformer experiments, we see that on Emo2Mix-Neutral the Sepformer-WJ2 model has a lower SDRi of 12.0 dB compared to 15.5 dB for Sepformer-L2M. The difference in SDRi performance for Emo2Mix-Strong is 1.9 dB on Sepformer-L2M compared to 2.6 dB for Sepformer-WJ2. A similar trend can be found for the ConvTasNet model, with the SDRi dropping to barely above noise at 0.9 dB on Emo2Mix strong. This mirrors the finding of~\cite{gauy2022pretrained} where emotional recognition performance benefited from non-emotional large scale pretraining.

\subsection{Performance Comparisons by emotions}
To streamline the performance comparison, we only report here the SI-SDRi metric of the best performing model, Sepformer-L2M. Table~\ref{tab:sep-l2m-nrml} shows the results of the model tested on the Emo2Mix-Normal dataset, while Table~\ref{tab:sep-l2m-strg} shows the results of the model tested on the Emo2Mix-Strong dataset. In each of the tables, the highest value in each row is listed in bold and while lowest value is underlined. 

In the first column of each table, we also report the comparison with the RAVDESS2Mix dataset (RD2M) to which consists only of a combination of 1 emotional utterance from the RAVDESS dataset and 1 neutral utterance from the Libri2Mix Test dataset. In this comparison we also see that there is a minimal variation in the RD2M dataset across the various emotions of 0.9dB for both the Strong and Normal emotional intensities. Meanwhile the average difference between the min and max emotions across combinations is 2.5dB for the normal intensity and 6.1dB for the strong intensity, with the lowest being 1.5dB for the ``sad" emotion row in the normal dataset. Across all combinations in the entire dataset, the difference between the min and max for Emo2Mix-normal is 4.7dB and for Emo2Mix-strong is 9.3dB.

Considering the results on the Emo2Mix-Normal dataset in Table~\ref{tab:sep-l2m-nrml} we see a trend that having an utterance with the surprise emotion consistently results in the worst performance compared to the other emotions, with the worst results arising from a combination of two surprise utterances. One possible explanation is that the actors in the dataset simply found it difficult to express surprise at a low emotional intensity and exhibited the most intense emotions in their surprised utterances compared to the other emotions expressed at normal intensity. Meanwhile sad and angry are emotions expressed at normal intensity could simply be close to neutral, resulting in better SI-SDRi results compared to the other emotions.

Finally, considering the results on the Emo2Mix-Strong dataset in Table~\ref{tab:sep-l2m-strg} we see a slightly different dynamic. The Calm utterances consistently have the best results in the Strong dataset compared to the Normal dataset. This could be that Calm utterances are closer to neutral at Strong emotional intensity and thus their contrast with the other strong emotions results makes it easier for the model to differentiate the speakers. 

\section{Conclusion}
In this work we have shown that, contrary to previous work~\cite{ravdess2mix}, emotions play an important role in the speech separation performance. Our results show that models trained on neutral speech will suffer a performance degradation when the mixture contains strong emotional expressions at inference time. Thus, this work makes the case for the need of including emotional speech into speech separation training datasets.

\section{Acknowledgements}
This research is supported by ST Engineering Mission Software \& Services Pte. Ltd under a collaboration programme (Research Collaboration No: REQ0149132). The computational work for this article was partially performed on resources of the National Supercomputing Centre, Singapore (https://www.nscc.sg).

\bibliographystyle{IEEEtran}
\bibliography{manuscript}

\end{document}